# Calculating Biological Behaviors of Epigenetic States in Phage λ Life Cycle

X.-M. Zhu[1], L. Yin[2], L. Hood[3], and P. Ao[4],*

[1] GENMATH, Corp. 5525 27th Ave.N.E., Seattle, WA 98105, USA
[2] School of Physics, Peking University, Beijing 100871, PR China
[3] Institute for Systems Biology, 1441 N. 34 St., Seattle, WA 98103, USA
[4] Departments of Mechanical Engineering and Physics, University of Washington, Seattle, WA 98195, USA

**Abstract**

The biology and behavior of bacteriophage λ regulation was the focus of classical investigations of molecular control of gene expression. Both qualitative and quantitative aspects of this behavior have been systematically characterized experimentally. Complete understanding of the robustness and stability of the genetic circuitry for the lysis-lysogeny switch remains an unsolved puzzle. It is an excellent test case for our understanding of biological behavior of an integrated network based on its physical, chemical, DNA, protein, and functional properties. We have used a new approach to non-linear dynamics to formulate a new mathematical model, performed a theoretical study on the phage λ life cycle, and solved the crucial part of this puzzle. We find a good quantitative agreement between the theoretical calculation and published experimental observations in the protein number levels, the lysis frequency in the lysogen culture, and the lysogenization frequency for the recently published mutants of $O_R$. We also predict the desired robustness for the λ genetic switch. We believe that this is the first successful example in the quantitative calculation of robustness and stability of phage λ regulatory network, one of the simplest and most well-studied regulatory systems.

……………………………..
* Corresponding author: Dr. Ping Ao, Department of Physics, University of Washington, Seattle, WA 98195, USA; Phone: (206) 543-3901; E-mail: ao@dirac.phys.washington.edu

## 1. Introduction

In order to understand the regulatory control of the phage λ life cycle (Ptashne, 1992; Ptashne and Gann, 2002; Little *et al.,* 1999) quantitative models (Ackers, *et al.*, 1982; Shea and Ackers, 1985; Reinitz and Vaisnys, 1990; Arkin *et al.*, 1998; Aurell and Sneppen, 2002) based on physical and chemical characteristics have been developed for λ $O_R$ regulatory networks. The objective has been to provide a single coherent model from which major experimental observations may be described: the development pathways leading to the two kinds of life cycles of λ, the maintenance of the lysogenic state, and the robustness of the epigenetic states. Modeling results so far, however, are unable to provide a quantitative agreement with experimental observations (Reinitz and Vaisnys, 1990; Aurell and Sneppen, 2002). With a new mathematical modeling approach, which integrates data from molecular levels of DNA-protein and protein-protein interactions, to the protein expressions and biological functions, we report here a successful quantitative connection from the individual molecular elements and the experimentally determined parameters to the characteristics of the epigenetic states: the protein numbers in each state, the fluctuations of protein numbers in each state, the intrinsic stability of the these states, and the changes of these qualities with mutations and experimental conditions. Our calculated results agree quantitatively with all the available experimental observations including the observed robustness of the phage λ epigenetic states. In addition to this agreement we are able to demonstrate how the "landscape" of epigenesis, e.g. the pathways and valleys (Reik and Dean, 2002), can be constructed mathematically from the physical and chemical model at the molecular level, which provides a visualization of the full potential of the network dynamics.

We first summarize the salient experimental and mathematical results on lambda phage. Upon infection of the host *E. coli* cell, the bacteriophage λ enters either of the two different life cycles. It can direct the cell to produce new λ phage particles, resulting in the lysis of the cell. Or, it can establish dormant residency in the lysogenic state, integrating its genome into the DNA of its host and replicate as a part of the chromosome. In these two different life cycles, different sets of genes are expressed. The genetic switch controlling the function of phage λ consists of two regulatory genes, cI and cro, and the regulatory regions, $O_R$ and $O_L$ on the λ DNA. Established lysogeny is maintained by the protein CI which blocks operators $O_R$ and $O_L$, and thereby prevents transcription of all lytic genes including cro (Ptashne, 1992; Ptashne and Gann, 2002). In lysogeny the CI number functions as an indicator of the state of the bacterium: If DNA is damaged such as by the UV light the protease activity of RecA is activated, leading to degradation of CI. A low CI number allows for transcription of the lytic genes, starting with cro, the product of which is the protein Cro.

According to the biological picture elegantly presented by Ptashne (Ptashne, 1992; Ptashne and Gann, 2002) and concisely summarized by Aurell *et al.* (Aurell *et al.*, 2002), the λ lysogeny maintenance switch is centered around operator $O_R$, and consists of three binding sites $O_{R1}$, $O_{R2}$ and $O_{R3}$, each of which can be occupied by either a Cro dimer or a CI dimer. These three binding sites control the activity of two promoters $P_{RM}$ and $P_R$ for respectively cI and cro transcriptions, illustrated in Fig.1. The transcription of cro starts at $P_R$, which partly overlaps $O_{R1}$ and $O_{R2}$. The transcription of cI starts at $P_{RM}$, which overlaps $O_{R3}$. The occupation of those sites hence prohibits the corresponding transcription of proteins. The affinity of RNA polymerase for the two promoters, and subsequent production of the two proteins, depends on the pattern of Cro and CI bound to the three operator sites and thereby establishes lysogeny with about 100--600 CI molecules per bacterium. If, however, CI number becomes sufficiently small, the increased production of Cro flips the switch to lysis.

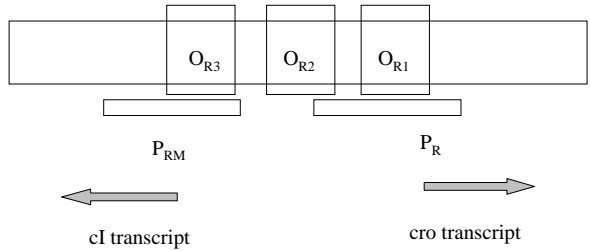

Figure 1. The phage λ regulatory region including promoters for the cI and cro genes and their overlapping operator sites. cI is transcribed when $O_{R3}$ is unoccupied and $O_{R2}$ is occupied by CI. cro is transcribed when both $O_{R2}$ and $O_{R1}$ are unoccupied.

The intrinsic stability of lysogenic state is measured by the amount of free phage in cultures of the lysogen. The recent three experiments by Aurell *et al.* (Aurell *et al.*, 2002), Rozanov *et al*. (Rozanov *et al.*, 1998), and Little *et al.* (Little *et al.*, 1999) on the frequency of spontaneous induction in strains deleted for the recA gene designed to avoid complications from the SOS system have generated the similar results, as reviewed by Aurell *et al.*. Among these consistent experimental data, the measurement by Little *et al.* appears to be well controlled and more systematic. We will take their results as our primary data for comparison. We do, however, take the note that there is an indication that the wild type may be more stable than suggested by those published data (Little, private communication), which posts an additional challenge to both experimentalists and modelers.

## 2. New Stochastic Nonlinear Dynamics Modeling

The minimal mathematical model which embodies Fig.1 is two coupled equations for the time rate of change of numbers of CI ($N_{CI}$) and Cro ($N_{Cro}$) in a cell (Shea and Ackers, 1985; Reinitz and Vaisnys, 1990), as systematically reviewed in Aurell *et al*. (Aurell *et al.*, 2002),

$$dN_{CI}/dt = f_{CI}(N_{CI}, N_{Cro}) - N_{CI}/\tau_{CI} + \zeta_{CI}(N_{CI}, N_{Cro}, t)$$

$$dN_{Cro}/dt = f_{Cro}(N_{CI}, N_{Cro}) - N_{Cro}/\tau_{Cro} + \zeta_{Cro}(N_{CI}, N_{Cro}, t) \qquad (1)$$

Here $f_{CI}$ and $f_{Cro}$ are the production rate for CI and Cro proteins in a host cell. $\tau_{CI}$ and $\tau_{Cro}$ are their effective decay constants. $f_{CI}$ and $f_{Cro}$ are functions of CI and Cro dimerization constants and the Gibbs free energies for their bindings to the three operator sites $O_{R1}$, $O_{R2}$ and $O_{R3}$ (see Appendix for details), the rate of initiation of transcription from $P_{RM}$ when stimulated by CI bound to $O_{R2}$, $T_{RM}$, and when not stimulated $T_{RM}^u$, and the efficiency of translation of the mRNA transcripts into protein molecules, the number of CI (Cro) molecules produced per transcript, $E_{cI}$ ($E_{cro}$).

The decay constant $\tau_{CI}$ is proportional to the bacterial life-time, since CI molecules are not actively degraded in lysogeny, while $\tau_{Cro}$ is about 30% smaller (Pakula *et al.*, 1986). We note that there is considerably more experimental uncertainty in the binding of Cro, both to other Cro molecules and to DNA, than the binding of CI, see e.g. Darling *et al.* (Darling *et al.*, 2000). In this minimal model of the switch, we take $\tau_{CI}$ and $\tau_{Cro}$ from data, and deduce $f_{CI}$ and $f_{Cro}$ at non-zero numbers of both CI and Cro with a standard set of assumed values of all binding constants, as done in Aurell *et al.* (Aurell *et al.*, 2002). Nevertheless, exactly numerical values of those parameters are not essential here. There is another important adjustment made below of the DNA-protein binding energies to take the *in vivo* and *in vitro* differences into account.

Random forces are represented by $\zeta_{CI}(N_{CI}, N_{Cro}, t)$ and $\zeta_{Cro}(N_{CI}, N_{Cro}, t)$ in Eq.(1). Their effects were not quantitatively studied in earlier mathematical modelings (Ackers, *et al.*, 1982; Shea and Ackers, 1985; Reinitz and Vaisnys, 1990). The numbers of CI and Cro are only in the range of hundreds, suggesting the importance of stochastic effects. The actual protein production process is also influenced by many chance events, such as the time it takes for a CI or a Cro in solution to find a free operator site, or the time it takes a RNA polymerase molecule to find and attach itself to an available promoter. As a minimal model of the switch with finite-N noise, we therefore consider two independent standard Gaussian and white noise sources to describe all possible chance effects (Little, private communication). For more mathematically oriented readers the noise should be understood as due to the Wiener process.

We comment that the protein binding free energies as reviewed in Aurell *et al.* (Aurell *et al.*, 2002) are taken from *in vitro* studies. The *in vivo* protein-DNA affinities could depend on the ions present in the buffer solutions as well as other factors. This difference must be taken into consideration in the theoretical calculation. The data quoted in Darling *et al.* (Darling *et al.*, 2000) was obtained at KCl concentration of 200mM, which resembles *in vivo* conditions. Therefore, though we expect a difference between the *in vivo* and *in vitro* data, it should not be too large (Ptashne, 1992).

Instead of a direct stochastic simulation using Eq.(1), such as performed by Arkin *et al.* (Arkin *et al.*, 1998) to calculate the lysogenization frequency, or of the abstract analytic method of Aurell and Sneppen (Aurell and Sneppen, 2002) to calculate the switching rate from lysogenic state to lytic state, we take a different mathematical approach which offers an intuitive and direct grasp of the stability problem. It has been demonstrated (Ao, 2002; Kwon *et al.*, 2003) that there exists a unique decomposition such that Eq.(1) can be transformed into the following form:

$$\sum_{j = CI, Cro} [S_{i,j}(N_{CI}, N_{Cro}) + A_{i,j}(N_{CI}, N_{Cro})] dN_j/dt = -\nabla_i U(N_{CI}, N_{Cro}) + \xi_i(N_{CI}, N_{Cro}, t) \qquad (2)$$

here i = CI, Cro, $\nabla_{CI} = \partial/\partial N_{CI}$, $\nabla_{Cro} = \partial/\partial N_{Cro}$ with the semi-positive definite symmetric 2×2 matrix *S*, the anti-symmetric 2×2 matrix *A*, the single-valued scalar function *U*. The connection between the random force $\xi$ and the matrix *S* is similar to that of $\zeta$ and *D* of Eq.(1): Therefore *S* is a diagonal matrix in the present situation. (see the Appendix.)

Eq.(2) suggests the analogy to the dynamics of a massless charge particle in the presence of a magnetic field, with the particle velocity d**N**/dt moving in the two dimensional space defined by the numbers of proteins, $N_{CI}$, $N_{Cro}$, a situation familiar in physics and chemistry. Here $\mathbf{N}^\tau = (N_{CI}, N_{Cro})$ is the transpose of vector **N**. We may interpret the semi-positive definite symmetric *S* matrix as the friction matrix, and the anti-symmetric matrix *A* as the result of a `magnetic' field. The friction matrix *S* describes the ability of the system approaching to equilibrium, similar to the degradation of the proteins. Hence it may be called the degradation matrix. Finally, in the present analogy the scalar function U takes the role of a potential energy function. The usefulness of such potential function has been discussed in dynamics of gene regulatory networks (Sagai and Wolynes, 2003).

The potential U may be interpreted as the dynamical landscape of the gene regulatory network (Fig. 2), a concept proposed long ago in developmental biology (Reik and Dean, 2002). Under appropriate conditions a phage $\lambda$ will see two minima and one saddle point in the potential energy landscape. Those two minima correspond to the lytic and lysogenic states. The positions of the potential minima give the most probable protein numbers for lytic and lysogenic states. The results are presented in Table I. The relative levels of CI level in the lysogen state are calculated using the wild type as the standard (100%). The relative Cro levels in lysis state are also calculated similarly. By varying the parameters in the model, we may also obtain information on how the protein numbers change when the experimental condition varies. For example, we find that when temperature is raised, both CI number for lysogenic state and Cro number for lytic state increase.

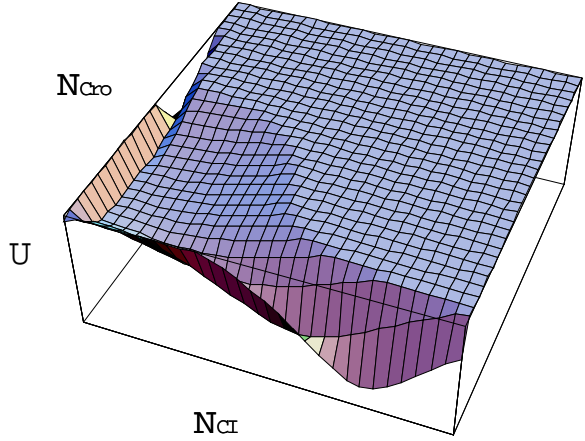

Figure 2. The potential U (plotted on log scale), the landscape of the phage λ regulatory network. The phage behaves essentially like a billiard ball in this landscape (more precisely, as a charged particle with negligible mass in a 'magnetic' field, see text). A straightforward interpretation of U is a measure of its implied confinement and paths. The protein numbers $N_{CI}$ and $N_{Cro}$ in a living cell are confined to the places where U is relatively low. If $N_{CI}$ and $N_{Cro}$ of a phage deviate from the minima of U, the phage will evolve towards one of the minima. There are two minima of U, corresponding to the lytic and lysogenic growth states, defining two attractive basins. While the phage $\lambda^+$ is at the lysogenic state, fluctuations in protein numbers $N_{CI}$ and $N_{Cro}$ may occasionally result in the phage moving to the lytic state. The shape of the potential determines that most likely, such a move will happen from the lysogenic minimum along the valley of the potential, into the lytic minimum. The highest point of the potential along the valley is the saddle point (position shown in Fig.3(a)). The potential difference between the saddle point and the lysogenic minimum is used to calculate the likelihood of such an event. Another major function of the potential U is to determine the evolution path for each phage. At the very early stage of infection (Ptashne, 1992), both CI and Cro are produced before turning on the phage's decision mechanisms. This phage will evolve to lytic or lysogenic state according to its position in the potential landscape.

The fluctuation of the protein number is determined by both the potential shape and the strength of degradation matrix *S*. The shape of the potential around its minima gives the CI and Cro distributions in lysogenic and lytic states. Such distributions have not yet been measured experimentally for phage λ so we do not include the prediction here. Occasionally, when the phage is at one of the minima, e.g. lysogenic state, a large fluctuation may bring it over to another minimum, lytic state. The probability is measured by its ability to go over the saddle point. The quantitative description is given by the Kramers rate formula (Kramers, 1940)

$$P = \omega_0 \exp(-U_b), \qquad (3)$$

with the barrier height $U_b = (U_{saddle} - U_{initial\ minimum})$, the difference in potential between the saddle point and the initial minimum, and the time scale associated with the attempt frequency $\omega_0$, determined by the degradation *S* and the curvatures around the saddle and the initial minimum. The dominant controlling factor is the barrier height $U_b$, showing an exponential sensitivity. The calculated fraction of lysogens that have switched to lytic state is presented in Table I.

## 3. Results

To calculate the lysogenization frequency requires additional information. Eq.(2) is a stochastic differential equation. Besides the equation itself, initial condition is needed to determine completely the behavior of the phage at the later stage. At the early stage of phage $\lambda$ infection before the decision mechanism is fully turned on (Ptashne, 1992), each phage acquires certain number of CI and Cro proteins. The CI and Cro protein numbers for different phage may be different. Therefore there is a distribution of CI and Cro numbers for an ensemble of phage. If this initial distribution is known, the lysogenization frequency is determined. However, we do not have such information. What we have is the outcome, the lysogenization frequency. Therefore we perform a 'reverse engineering' to determine the initial protein number from the outcome. This differs from Arkin *et al.* (Arkin *et al.*, 1998) who demonstrated the lysogenization may be calculated based on master equations for wild type under certain condition. For simplicity, we assume the initial protein distribution is uniform. We find that with the same initial protein distribution for all the different mutants growing under the same condition, measured in the relative CI level (with respect to CI in lysogen) and the relative Cro level (with respect to Cro in lysis), we can reasonably reproduce experimental results numerically with same qualitative trend.

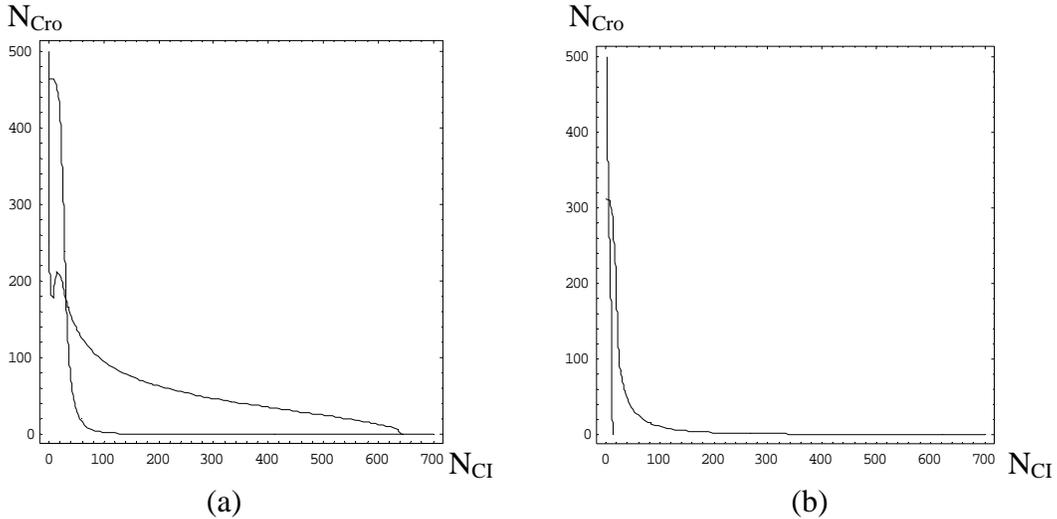

Figure 3. Lines $d\langle N_{CI}\rangle/dt = 0$ and $d\langle N_{Cro}\rangle/dt = 0$ for (a): wild type phage $\lambda^+(\lambda O_R 321)$; (b): mutant $\lambda O_R 123$. Here the average $\langle \ \rangle$ is performed over the noise. By definition, the position of the potential minima and saddle point coincide with the intersections of nullclines $d\langle N_{CI}\rangle/dt = 0$ and $d\langle N_{Cro}\rangle/dt = 0$. In the first three cases, there are three points where these two lines intersect. These three fixed points correspond to two potential (U) minima and one saddle point. The two minima represent lytic and lysogenic growth. Mutant $\lambda O_R 121$ and mutant $\lambda O_R 323$ have the similar three intersecting points. Therefore these strains have a switch structure. Comparison of the positions for the potential minima gives the relative CI level in lysogen and the relative Cro level in lysis listed in Table I. The last mutant ($\lambda O_R 123$) does not have such a switch structure. There is only one fixed point, corresponding to the lytic growth, as observed by Little *et al.* (Little *et al.*, 1999).

The mutant λ$O_R$3'23' measured by Little *et al.* (Little *et al.*, 1999) was first characterized by Hochschild *et al.* (Hochschild *et al.*, 1986). There is no quantitative measurement on its binding free energies yet. In order to produce the desired protein level, we find that the binding energy between $O_R$3' and Cro protein is 1.8 kcal/mol smaller than that of the $O_R$3 and Cro protein, consistent with the qualitative observation of Hochschild *et al.* We find slightly increased, 1 kcal/mol, of CI binding energy from $O_R$3 to $O_R$3', also consistent with their observation.

| Phage | Lysogenization frequency | Relative CI level in lysogen | Relative Cro level in lysis | Switching rate of lysogen to lytic state |
|---|---|---|---|---|
|  | Theoretical (experimental) | Theoretical (experimental) | Theoretical | Theoretical (experimental) |
| λ$^+$ | 90%  (63%) | 100%  (100%) | 100% | $2\times10^{-7}$  ($4\times10^{-7}$) |
| λ$O_R$121 | 70%  (57%) | 20%  (25-30%) | 100% | $2\times10^{-6}$  ($3\times10^{-6}$) |
| λ$O_R$323 | 10%  (33%) | 70%  (60-75%) | 70% | $5\times10^{-5}$  ($2\times10^{-5}$) |
| λ$O_R$3'23' | 80%  (60%) | 50%  (50-60%) | 130% | $5\times10^{-7}$  ($5\times10^{-7}$) |

Table I. The comparison between the theoretical and experimental results. The experimental results of Little *et al.* (Little *et al.*, 1999) are in the parentheses. The wild type data are used to fix parameters in our minimal model. Others are then calculated accordingly. The bi-stability may be realized by suppressing the lyses, the so-called anti-immune phenotype (Calef *et al.*, 1971; Eisen *et al.*, 1970). The relative Cro level predictions can be tested in this way. Relative protein levels are calculated against their wild type values respectively: 644 for CI and 464 for Cro. The unit of the switching rate to lytic is per minute.

Another natural consequence of this mathematical model is the robustness of the λ switch. Living organisms survive various sources of impact and are highly robust to certain fluctuations. Simultaneously, these organisms have certain tolerance to genetic polymorphism. Both of these properties are presumably the consequences of evolution. Robustness is understood as the ability of keeping the systems behavior intact while the details may change significantly. The genetic stability of the λ switch is inferred from Table I. Little *et al.* (Little *et al.*, 1999) have given an elaborated discussion from experimental point of view and concluded that qualitative pattern of λ gene regulation persists for the mutations in their experiments. The theoretical results by Aurell *et al.* (Aurell *et al.*, 2002) were unable to establish the stabilities for the some of the mutants listed in Table I. Our calculation reproduces Little *et al.*'s experimental results and therefore explains their observations from physical and chemical first principles.

The insensitivity of the stability to smooth parameter changes from our analysis agrees with the following theoretical observations in Aurell *et al.* (Aurell *et al.*, 2002). The switch is in general stable though smaller Cro initialization rate $T_R$ makes the switch more stable. Aurell *et al.* varied Cro degradation rate $\tau_{Cro}^{-1}$, $T_R$, and finally $T_R$ and $E_{cro}$ by keeping the product $T_R \times E_{cro}$ fixed. Thus, with given binding affinities the mean rate of switching increases with any of $E_{cro}$, $T_R$ and $\tau_{Cro}$, but by varying two of these parameters in opposite directions, they reproduced the measured stability. CI influences the stability of a lysogen by binding to $O_R1$ and $O_R2$, which determines the fraction of time $P_R$ is open. CI also indirectly influences the stability of a lysogen through the binding to $O_R3$, which determines the fraction of time $P_{RM}$ is open. Any CI affinity increase could be compensated by a corresponding increase in $T_R$.

Viewing the existence of two minima and one saddle point as the hallmark for the stability of the switch, we further test of the stability of the wild type based by varying parameters. We find that for the CI dimer number varying from $2\times10^{-2}$ to $1\times10^2$ relative to the original value and for Cro dimer number from $5\times10^{-4}$ to $1\times10^4$ relative to the original value, the switch is stable. We also let the CI degradation time $\tau_{CI}$ vary from 0.1 to 15 relative to its original value and $\tau_{Cro}$ vary from $1\times10^{-2}$ to $5\times10^2$ relative to its original value, again the switch is stable. Note that one of parameters is varied over 8 orders of magnitude. With all the above tests on the stability, we conclude that this phage λ switch is extremely stable, insensitive to changes in parameters.

## 4. Discussion

One may wonder why previous mathematical modeling failed to explain the robustness of the lysogenic state of phage λ. In the light of present quantitative modeling, two reasons may be offered. First, the difference between the *in vivo* and *in vitro* molecular parameters, though small, does play a dominant role. Both the effect of chemical environment on binding free energies and the effect of cooperativity among proteins in the living organism are included in this difference. This small difference has been difficult to measure. Second, previous methods, such as that of Aurell and Sneppen (Aurell and Sneppen, 2002), are rigid. No additional stochastic processes rather than those dictated by the assumed chemical reaction rates have been considered in those methods.

Our analysis demonstrates that both qualitative and quantitative agreements between theoretical calculations and the experimental data can be established in an integrative modeling. Our model offers an intuitively simple quantification of the robustness and stability of the phage λ genetic switch: the potential landscape for the network dynamics, which may be useful in tackling more complex gene regulatory networks.

## Acknowledgements

We thank G.K. Ackers, D. Galas, and J.W. Little for valuable comments and critical discussions. This work was support in part by the Institute for Systems Biology (P.A. and L.H.) by USA NIH grant under HG002894-01 (P.A.) and by USA NSF grant under DMR 0201948 (L.Y.).

**Appendix**

**I. Details for Eq.(1):**

$$dN_{CI}/dt = f_{CI}(N_{CI}, N_{Cro}) - N_{CI}/\tau_{CI} + \zeta_{CI}(N_{CI}, N_{Cro}, t)$$
$$dN_{Cro}/dt = f_{Cro}(N_{CI}, N_{Cro}) - N_{Cro}/\tau_{Cro} + \zeta_{Cro}(N_{CI}, N_{Cro}, t) \quad (1)$$

Following Ackers *et al.* (Ackers *et al.*,1982) and Aurell *et al.* (Aurell *et al.*, 2002), we encode a state *s* of CI and/or Cro bound to $O_R$ by three numbers (i,j,k) referring to $O_{R3}$, $O_{R2}$ and $O_{R1}$ respectively. The coding is 0 if the corresponding site is free, 1 if the site is occupied by a CI dimer, 2 if the site is occupied by a Cro dimer, and 3 if the site is occupied by an RNAp. The probability of a state *s* with $i_s$ CI dimers and $j_s$ Cro dimers bound to $O_R$ is in the grand canonical approach of Shea and Ackers (Shea and Ackers, 1985)

$$P_R(s) = Z^{-1} [CI]^{i_s} [Cro]^{j_s} [RNAp]^{k_s} \exp(-G(s)/T) \quad (I.1)$$

For example, if CI occupies $O_{R1}$, Cro $O_{R2}$ and $O_{R3}$, we have $i_s = 1$, $j_s = 2$, $k_s = 0$, and $P_R(s) = P_R(221)$. There are total 40 states represented by *s*. The normalization constant $Z$ is determined by summing over *s*: $Z = \Sigma_s [CI]^i_s [Cro]^j_s [RNAp]^k_s \exp(-G(s)/T)$.
The converting factor between protein number and concentration inside the bacterium is estimated to be $1.5 \times 10^{-11}$.

RNAp occupies either $O_{R1}$ and $O_{R2}$, or $O_{R2}$ and $O_{R3}$, not other configurations. We further simplify the expression of $P_R(s)$ by noticing that the controlling of $O_R$ is operated by CI and Cro proteins, not RNAp (Ptashne, 1992). If $O_{R1}$ and $O_{R2}$ is unoccupied by either CI or Cro, RNAp binds to them with a probability determined by RNAp binding energy. The case that RNAp first binds to $O_{R1}$ and $O_{R2}$, then blocking the CI and Cro binding is excluded based on the assumption that only CI and Cro controls the regulatory behavior. In addition to experimental observation, this assumption is justifiable if the time scale associated with CI and Cro binding is shorter than the RNAp binding. Except for an overall constant, which we include into the rate of transcription, the RNAp binding is no longer relevant. We therefore take it out of the expression $P_R(s)$ and drop the R subscript. The total number of states is reduced to 27. This simplification was first used by Aurell and Sneppen (Aurell and Sneppen, 2002).

The dimer and monomer numbers are determined by the formation and de-association of dimers, which gives the relation of dimer concentration to the total concentration of proteins as

$$[CI] = [N_{CI}]/2 + \exp(\Delta G_{CI}/RT)/8$$
$$+ ([N_{CI}] \exp(\Delta G_{CI}/RT)/8 + \exp(2\Delta G_{CI}/RT)/64]^{1/2} \quad (I.2)$$

Here $\Delta G_{CI} = -11.1$ kcal/mol is the dimer association free energy for CI.
Similar expression for [Cro] is

$$[Cro] = [N_{Cro}]/2 + \exp(\Delta G_{cro}/RT)/8$$
$$+ ([N_{Cro}]\exp(\Delta G_{Cro}T)/8 + \exp(2\Delta G_{Cro}T)/64\}]^{1/2} \quad (I.3)$$

Here $\Delta G_{Cro} = -7$ kcal/mol is the dimer association free energy for Cro.

CI and Cro are produced from mRNA transcripts of cI and cro, which are initiated from promoter sites $P_{RM}$ and $P_R$. The rate of initiation of transcription from $P_{RM}$ when stimulated by CI bound to $O_{R2}$ is denoted $T_{RM}$, and when not stimulated $T_{RM}^u$. The number of CI molecules produced per transcript is $E_{cI}$. The overall expected rate of CI production is

$$f_{CI}(N_{CI}, N_{Cro}) = T_{RM} E_{cI} [P(010) + P(011) + P(012)] + T_{RM}^u E_{cI} [P(000)$$
$$+ P(001) + P(002) + P(020) + P(021) + P(022)]. \quad (I.4)$$

Similarly, the overall expected rate of Cro production is

$$f_{Cro}(N_{CI}, N_{Cro}) = T_R E_{cro} [P(000) + P(100) + P(200)]. \quad (I.5)$$

Putting together the production and decay of proteins, we obtain coupled equations for the time rate of change of numbers of CI and Cro in a cell (Reinitz and Vaisnys, 1990)

$$dN_{CI}/dt = f_{CI}(N_{CI}, N_{Cro}) - N_{CI}/\tau_{CI}$$
$$dN_{Cro}/dt = f_{Cro}(N_{CI}, N_{Cro}) - N_{Cro}/\tau_{Cro}. \quad (I.6)$$

The need to include fluctuations has been emphatically discussed in recent publications (Arkin *et al.*, 1998; Aurell *et al.*, 2002).

**II. Parameters:**

We use basically the same set of parameters as listed in Table I as that of Aurell and Sneppen (Aurell and Sneppen, 2002). Those values were taken from various sources (Koblan and Ackers, 1991 and 1992; Takeda *et al.*, 1989 and 1992; Kim *et al.*, 1987; Jana *et al.*, 1997; Darling *et al.*, 2000).

$RT = 0.617$ kcal/mol,
Effective bacterial volume $= 0.7 \times 10^{-15}$ *l*,
  $E_{cro} = 20$,
  $E_{cI} = 1$,
  $T_{RM} = 0.115$ /s,
  $T_{RM}^u = 0.01045$ /s,
  $T_R = 0.30$/s,
  $\tau_{CI} = 2943$ s,
  $\tau_{Cro} = 5194$ s,
*in vitro* free energy differences:

ΔG (001) = -12.5 kcal/mol,
ΔG (010) = -10.5 kcal/mol,
ΔG (100) = -9.5 kcal/mol,
ΔG (011) = -25.7 kcal/mol,
ΔG (110) = -22.0 kcal/mol,
ΔG (111) = -35.4 kcal/mol,
ΔG (002) = -14.4 kcal/mol,
ΔG (020) = -13.1 kcal/mol,
ΔG (200) = -15.5 kcal/mol,

*in vitro* free energy differences for Hochschild mutant:
ΔG (100) = -10.5 kcal/mol,
ΔG (200) = -13.7 kcal/mol,

*in vivo* free energy differences = *in vitro* free energy differences + ΔG. This ΔG is –2.5 kcal/mol for CI, -4.0 kcal/mol for Cro as found in our numerical calculations. Those two variations between *in vivo* and *in vitro* free energy difference is less than 30%. The ΔG for cooperation energy is –3.7 kcal/mol, comparing with the *in virtue* value of –2.7 kcal/mol. This relatively large difference may be due to the looping effect (Dodd *et al.*, 2001) implicitly considered in our present modeling.

## III. Decomposition leading from Eq.(1) to Eq.(2).

Because the mathematical method we use to analyze the stochastic equation Eq.(1) is new, we will first describe briefly its essence. The method provides a visualization of the network dynamics. Its equivalence has been widely used in physics and chemistry. We start by rewriting Eq.(1) so that its notation is more familiar. Defining a vector $\mathbf{r} = (x,y)$, $x=N_{CI}$, $y=N_{Cro}$, Eq.(1) becomes

$$d\mathbf{r}/dt = \mathbf{F}(\mathbf{r}) + \zeta(\mathbf{r},t) \tag{III.1}$$

with $F_x=f_{CI}(\mathbf{r}) - x/\tau_{CI}$, $F_y=f_{Cro}(\mathbf{r}) - y/\tau_{Cro}$, $\zeta_x=\zeta_{CI}(\mathbf{r},t)$ and $\zeta_y=\zeta_{Cro}(\mathbf{r},t)$. Eq.(III.1) is supplemented by the relations

$$\begin{aligned}&<\zeta(\mathbf{r},t)>=0,\\&<\zeta_x(\mathbf{r}\ t)\zeta_x(\mathbf{r},t')> = 2\ D_x(\mathbf{r})\ \delta(t-t'),\\&<\zeta_y(\mathbf{r},t)\zeta_y(\mathbf{r},t')> = 2\ D_y(\mathbf{r})\ \delta(t-t'),\\&<\zeta_x(\mathbf{r},t)\ \zeta_y(\mathbf{r}\ t')> = 0\end{aligned} \tag{III.2}$$

which define a diagonal diffusion matrix D. Eq.(III.1) corresponds to the dynamics of a particle moving in two dimensional space with both deterministic and random forces. It is easy to check that $\nabla\bullet\mathbf{F}(\mathbf{r}) \neq 0$ and $\nabla\times\mathbf{F}(\mathbf{r}) \neq 0$ in general. Therefore $\mathbf{F}(\mathbf{r})$ cannot be represented by the gradient of a scalar potential. Recall that the simplest case in two dimensional motion when $\nabla\bullet\mathbf{F}(\mathbf{r}) \neq 0$ and $\nabla\times\mathbf{F}(\mathbf{r}) \neq 0$ is a charged particle moving in a magnetic field, Eq.(III.1) may be equivalent to

$$\eta \, d\mathbf{r}/dt + \mathbf{B}(\mathbf{r}) \times d\mathbf{r}/dt = -\nabla U(\mathbf{r}) + \xi(\mathbf{r},t) \tag{III.3}$$

with $\mathbf{B}(\mathbf{r})$ the magnetic field, $\eta$ the degradation matrix in the present paper (corresponding to friction matrix), and the random or stochastic force $\xi(\mathbf{r},t)$ which is related to the diagonal degradation matrix $\eta$,

$$\begin{aligned}
&<\xi(\mathbf{r},t)> = 0 \\
&<\xi_x(\mathbf{r},t)\, \xi_x(\mathbf{r},t')> = 2\, \eta_x\, \delta(t-t') \\
&<\xi_y(\mathbf{r},t)\, \xi_y(\mathbf{r},t')> = 2\, \eta_y\, \delta(t-t') \\
&<\xi_x(\mathbf{r},t)\, \xi_y(\mathbf{r},t')> = 0
\end{aligned} \tag{III.4}$$

Eq.(III.3) is the same as Eq.(2) in the main text when $S_{i,j} = \eta_{i,j}$, $A_{i,j} = B\, \varepsilon_{i,j}$, here $\varepsilon_{i,j} = -\varepsilon_{j,i}$, $i,j = x,y$.

It is easy to see that Eq.(III.1) can be derived from Eq.(III.3). However, deriving Eq.(III.3) from Eq.(III.1) is highly non-trivial. The procedure is described in Ao (2002) and Kwon *et al.*(2003). We briefly summarize their conclusions here.
  i) Under general conditions, Eq.(III.3) can be derived from Eq.(III.1);
  ii) The connection between Eq.(III.3) and Eq.(III.1) is unique.

**IV. Intermediate results and parameters.**

In Table I, we have presented the final results. There are some intermediate results together with the parameters of interest.

The degradation constants: $\eta_{11} = 0.056 \times \tau_{CI}/N_{CI}(\text{lysogen})$, $\eta_{22} = 0.056 \times \tau_{Cro}/N_{Cro}(\text{lysis})$, here $N_{CI}(\text{lysogen})$ is the $N_{CI}$ value at the lysogen point and $N_{Cro}(\text{lysis})$ the $N_{Cro}$ value at the lysis point :

| | | |
|---|---|---|
| $\lambda^+$ | $\eta_{11} = 0.254$ | $\eta_{22} = 0.634$ |
| $\lambda O_R 121$ | $\eta_{11} = 0.254/20\%$ | $\eta_{22} = 0.634/100\%$ |
| $\lambda O_R 323$ | $\eta_{11} = 0.254/70\%$ | $\eta_{22} = 0.634/70\%$ |
| $\lambda O_R 3'23'$ | $\eta_{11} = 0.254/50\%$ | $\eta_{22} = 0.634/130\%$ |

here 20% for $\lambda O_R 121$ and so on are the relative CI(Cro) level in lysogen(lysis).

We assume degradation matrix is a diagonal constant matrix. Similar to Aurell and Sneppen (Aurell and Sneppen, 2002), we assume the fluctuations in Eq.(1) scale with the square root of protein number divided by relaxation time. This leads to the relation that degradation scales inversely with the protein number and proportional to relaxation time, according to the Einstein relation (Ao, 2002). The overall constant 0.056 is a fit to the experiment.

Barrier height from the lysogen minima to the saddle points:

$\lambda^+$        $U_b = 7.49$

$\lambda O_R 121$    $U_b = 5.14$

$\lambda O_R 323$    $U_b = 2.14$

$\lambda O_R 3'23'$  $U_b = 6.63$

Attempt frequency:
$$\omega_0 = 4.0 \times 10^{-4} / \text{minute}$$

The lysogenization frequency is estimated assuming that at the early state of phage $\lambda$ infection, before the decision mechanism is fully turned on, each phage acquires certain number of CI and Cro proteins. The CI and Cro protein numbers for different phage maybe different. The distribution of CI and Cro numbers for an ensemble of phage is assumed to be uniform. The distribution for the wild type is the inside area of a circle centered at $N_{CI}=50$, $N_{Cro}=0$,

$$(N_{CI})^2 + (N_{Cro})^2 = 150^2 \qquad N_{CI}>0, N_{Cro}>0. \qquad (IV.1)$$

For different mutants growing under the same condition, we assume that the perimeter of such a distribution scale with the CI level in lysogen and the Cro level in lysis. For example, for $\lambda O_R 121$, the ellipse of the distribution is given by

$$[(N_{CI})/20\%]^2 + [(N_{Cro})/100\%]^2 = 150^2 \qquad N_{CI}>0, N_{Cro}>0, \qquad (IV.2)$$

here 20% and 100% are the relative CI level in lysogen and relative Cro level in lysis compared with the wild type.

intrinsic, in equal footing: Our assumption that all are represented by the Gaussian white noise may be an oversimplified one. For example, some probability events, such as the prm240 mutation which may be equivalent to destroy the gene switch, as suggested by Little unpublished data, could have totally different biological consequence therefore may not be well approximated by the Gaussian white noise in the present model. The neglecting such large, rare, and destructive events indeed enhances the gene switch stability against small and frequent Gaussian white noise. Particularly it may greatly reduce the wild type switching rate. We found that in such cases, other molecular parameters, such as the binding free energies, are nevertheless less sensitive to those rare events. Same overall good agreement with experimental data can be obtained. We will discuss this feature elsewhere in detail.